\documentclass[11pt]{JHEP3}
\usepackage{epsfig}
\usepackage{times}
\newcommand{\nf}{x_{\rm \scriptscriptstyle H}}
\def\la{\lower.5ex\hbox{$\; \buildrel < \over \sim \;$}}
\def\ga{\lower.5ex\hbox{$\; \buildrel > \over \sim \;$}}
\def\apj{ApJ}

\def\apjl{ApJL}

\title{Primordial magnetic fields and the HI signal from the epoch of reionization}
\author{Shiv K. Sethi \\ 
Raman Research Institute, Sadashivanagar, Bangalore 560080, India}
\author{ Kandaswamy Subramanian \\
 Inter-University Center for Astronomy \& Astrophysics, Post Bag 4, Ganeshkhind, Pune 411007, India\\
}

\vspace{+0.4cm}

\abstract{The implication of primordial magnetic-field-induced  structure formation for the HI
signal from the epoch of reionization is studied. Using semi-analytic models, we compute both
the  density and ionization inhomogeneities in this scenario.
We show that: (a) The global HI signal can only be seen in emission, unlike
in the standard $\Lambda$CDM models, (b) the density perturbations induced by primordial fields,
leave distinctive signatures of the magnetic field Jeans' length   
on the HI two-point correlation function, (c) the length
scale of ionization inhomogeneities is $\la 1 \, \rm Mpc$. 
We find that the peak expected
signal (two-point correlation function) 
 is $\simeq 10^{-4} \, \rm K^2$ in the range of scales
 $0.5\hbox{--}3 \, \rm Mpc$
for  magnetic field strength in the range
$5 \times 10^{-10} \hbox{--}3 \times 10^{-9} \, \rm G$. We also discuss the detectability of the
HI signal. The   angular resolution of the
on-going and planned radio interferometers
allows one to probe only
the largest
magnetic field
strengths that we consider.
They   have
the sensitivity to detect the  magnetic field-induced features. We show that the
future SKA has both
the angular resolution and the sensitivity to detect the magnetic field-induced signal in the entire range of magnetic field values we consider, in an integration time  of one week.}

\begin{document}

\section{Introduction}
Understanding the epoch of reionization is one of the outstanding challenges of 
modern cosmology. In recent years, a partial understanding  of this 
important  stage in the history of the universe has emerged. Gunn-Peterson tests
have revealed the existence of neutral hydrogen (HI)  at redshifts $z \ga 5.7$ (\cite{fan,white,djorgo,fan02,becker}). The detection
of temperature-polarization cross-correlation and the polarization-polarization correlation
at large angular scales ($\ell \la 10$) in the WMAP data have given firm evidence that
the universe reionized 
 around
$z \simeq 10$ (e.g. \cite{wmap}). 

The most direct approach to studying this epoch is to attempt to detect the redshifted 
21-cm line of the neutral hydrogen during the period when   the universe makes a transition from being fully
neutral to fully ionized. There is substantial on-going effort in that direction and 
this also remains one of the primary goals of upcoming and 
future radio interferometers (e.g.  MWA, for a detail description see 
\cite{bmh06}, LOFAR \footnote{\tt www.lofar.org}, and  SKA
\footnote{\tt www.skatelescope.org/pages/page\_sciencegen.htm}). 
Primordial tangled magnetic fields can substantially alter the scenario of structure
formation of the universe (\cite{ss05}, hereafter Paper I
and references therein). 
It has been shown that magnetic field-induced structure formation 
can cause early reionization (Paper I, \cite{ts06a})
and might leave detectable signatures in the HI signal 
from the epoch of reionization (\cite{ts06b,sbk}).

In this paper, we study the implications of the existence of the primordial magnetic fields
for possible studies of the reionization epoch. We compute both the global HI signal
and its fluctuating component. In particular, we study in detail the nature of 
ionization inhomogeneities in the presence of magnetic fields, as their precise understanding is needed to correctly set the scale of the effect.   The detailed quantification  of this aspect constitutes the main difference 
between our present work and earlier studies. We also discuss in detail the detectability of the magnetic field-induced HI signal.

In \S 2, we briefly review the salient features of the structure formation process in the 
presence of primordial tangled magnetic fields. In 
%section 3
\S 3, we give results for the reionization
process. In \S 4, we compute and discuss the global HI signal. In \S 5 and 6, the fluctuating component of the HI signal is derived and discussed. The detectability of the 
resultant signal is also discussed. Section 7 is reserved for summary and conclusions.
 In this paper we
use  the parameters of the   FRW  model as favoured by the recent 
 WMAP results \cite{wmap}:
spatially flat FRW model 
with $\Omega_m = 0.3$ and $\Omega_\Lambda = 0.7$
 with  $\Omega_b h^2 = 0.022$  and
$h = 0.7$ \cite{freedman}.

\section{Magnetic fields in the post-recombination era}
Sethi and  Subramanian (2005) \cite{ss05}   analysed two possible effects of the presence of 
primordial magnetic fields in the post-recombination era: (a) magnetic fields
can induce formation of early structures, (b) the dissipation of magnetic field
energy owing to ambipolar diffusion and decaying turbulence can significantly alter
the thermal and ionization history. 

 The main difference between the standard $\Lambda$CDM models
and the magnetic field-aided structure
formation is that, in the latter case, the magnetic field induces additional 
matter density fluctuations whose 
%the matter  
power spectrum is given by: 
$P(k) \propto  k^n$ for $k \le k_J$. Here
$k_J \simeq 15 (10^{-9}\, {\rm G}/B_0) \, \rm Mpc^{-1}$ is the magnetic field 
Jeans' wave number.  The matter power spectrum  spectral index $n$ tends  to one as the magnetic field spectral index approaches 
$-3$ (for details and discussion see e.g. \cite{gs03,ss05}). Throughout this paper we use magnetic field power spectrum index $ -2.9$, which 
gives $n \simeq 1.1$.

This results in more power at smaller scales as compared to the standard $\rm \Lambda CDM$ 
 case.  In Paper I \cite{ss05}, we  showed that the 1$\sigma$ collapse
of structures is possible at large redshifts; the collapse redshift 
of  dark matter haloes is sensitive to the spectral index 
of the magnetic field power spectrum even
though it is fairly independent of the magnetic field strength. It was also shown that 
the mass dispersion $\sigma(R) \propto 1/R^2$ for scales above the Jeans' mass scales, which
is a much sharper fall than the usual case. This means that most of the power 
is concentrated 
around scales close to magnetic Jeans' scales. These considerations also give 
strong indication that the only allowed tangled primordial field models are the 
ones for which $n \simeq -3$ (\cite{ss05,gs03}). 

The magnetic field dissipation in the post-recombination era 
 can
also significantly
alter the thermal Jeans' length scale, by heating the matter up to $T_m \simeq 10^4 \, \rm K$, depending on the value of the magnetic field strength. Therefore, in this
scenario,  there is a play-off
between early structure formation owing to enhanced power at small 
scales and suppression of power owing to 
magnetic field Jeans' scales and the enhanced thermal Jeans' scale
 (for details see \cite{sns}). Taking into account 
all of these effects, the interesting range of 
magnetic field strength is: $3 \times 10^{-10} \, {\rm G} \le B_0 \le 3\times 10^{-9}\, {\rm G}$. 
 We study here the effects of primordial magnetic fields which are in the above
range.

\section{Reionization}
We model 
%the evolution of 
the reionization of the universe as described
in Sethi (2005) \cite{sethi05}(for more details see \cite{hh03}). In this
model, an HII sphere is carved around collapsed  haloes. 
%The reionization
Reionization
proceeds as more sources are born and as the radius of the HII region
around each source increases. 
%The reionization 
It is completed when the 
HII regions coalesce. In this scenario, the ionized fraction at any redshift
is given by:
\begin{equation}
f_{\rm ion}(z) = {4 \pi \over 3}\int_{0}^z dz' \int dM {d^2n \over dM dz'}(M,z') R^3(M,z,z')
\label{eq:ionfra}
\end{equation} 
This expression remains valid until $f_{\rm ion} \simeq 1$. 
Here the mass function $dn/dM$ is computed using the Press-Schechter formalism. 
 The radius of the HII region $R$ is computed by following 
its  evolution around a source with a 
given photon luminosity $\dot N_\gamma$ (in $sec^{-1}$) which is assumed to 
grow  linearly with the halo mass; we  fix the fiducial value of the photon
luminosity to be $\dot N_\gamma(0)$ 
at the mass scale
 $M = 5 \times 10^7 \, \rm M_\odot$ (see e.g. \cite{sethi05}). The radius of 
the evolving HII region is given by (e.g. \cite{wl04,sethi05}): 
\begin{equation}
{dR \over dt} -H R={(\dot N_\gamma - (4\pi/3) R^3 \alpha_B C  n_b^2 x_{\rm \scriptscriptstyle HI})  \over (\dot N_\gamma + 4 \pi R^2 x_{\rm \scriptscriptstyle HI} n_b)}
\label{stromsp}
\end{equation}
Here $\alpha_B$ is the case B recombination coefficient, $n_b$ is the 
background density of  baryons (excluding the helium atoms), and $C$ is the 
clumping factor.
We further assume the photon luminosity to be exponentially suppressed with time
constant of one-tenth of the Hubble time at the redshift at which the source is born. 
 The other parameter
used in the evolution of the HII region around each source is the average clumping 
factor $C$; we assume $C=2$ 
in the entire redshift range.

As noted above, the mass function of objects in
 the magnetic field-aided collapse of structures 
is expected to have much larger number of objects close to the magnetic Jeans' scale as 
compared to the usual case, with this number falling sharply at larger scales. 
Tashiro \& Sugiyama (2006a) \cite{ts06a} computed the reionization history in this case and showed that
 reionization is possible with star-formation efficiency markedly lower than
 in the usual case. We re-look at this issue within the framework of the  semi-analytic approach
described above, which explicitly takes into account the growth of HII bubbles around  sources. 

We show our results in Figure~1. In the Figure, we plot reionization history for two
values of magnetic field strength and the usual case 
 with zero magnetic fields.  
We adopt $\dot N_\gamma(0) = \{4 \times 10^{48}, 10^{49} \} \, \rm sec^{-1}$ 
for the two magnetic field cases shown in the figure, 
respectively with $B_0 = 3 \times 10^{-9} \, {\rm G}$ and 
$B_0 = 10^{-9}\, {\rm G}$.
% shown with lower value of photon luminosity for higher value 
%of magnetic field strength. 
For the standard case with zero magnetic field we take
$\dot N_\gamma(0) = 2.5 \times 10^{50}$.
% in the standard  case.  
All the reionization histories are
normalized to the WMAP result $\tau_{\rm reion} = 0.1$. \footnote{We note that  $\tau_{\rm reion}$ is the  integrated value between $30 \ge z \ge 0$, to be relevant for the 
WMAP results. Owing to magnetic field dissipation, partial reionization can lead
to a much larger 'reionization' optical depth at much higher redshifts. However, WMAP
results are not sensitive to such early partial ionization (Paper I)}

Photon luminosity can be cast in terms of the star formation efficiency, $f_{\rm eff}$ and escape fraction of hydrogen-ionizing photons $f_{\rm esc}$ from star-forming haloes (for more details see \cite{bl01}, p57). 
 A photon luminosity of $\dot N_\gamma(0) = 2 \times 10^{50}$ corresponds 
roughly to $f_{\rm esc}f_{\rm eff} = 0.01$ if the photons are emitted over a period 
one tenth of the Hubble time at $z \simeq 10$, using Scalo IMF. This means that reionization in the 
standard  zero field 
case requires both the star formation efficiency and escape fraction of roughly
$10\%$. The product of these two parameters could be  nearly two orders of magnitude smaller
if $B_0 \simeq 3 \times 10^{-9} \, \rm G$.  Thus one
needs a much lower efficiency of star formation for  models with primordial
magnetic fields. As both these parameters are highly uncertain
our analysis  doesn't allow us to constrain the magnetic field models 
from WMAP results alone (e.g. \cite{bl01}). 
However, we note here that  future
meaningful constraints on the star formation efficiency and escape fraction
could be used to rule/bear  out magnetic field-induced reionization.

 We discuss briefly each of the case of magnetic field-induced reionization:

$B_0 \la  5 \times 10^{-10} \, \rm G$: In this case, the objects that cause reionization are
 in the mass range $\le  \hbox{a few } 10^7 \, \rm M_\odot$ or these 
are  molecular-cooled haloes (for a review see e.g. \cite{bl01}). 
There are many aspects to the  nature of reionization caused by these objects that deserve 
further attention: (a) the temperature of the regions of the universe that are already 
(photo)ionized is $\simeq 10^4 \, \rm K$; this results in a thermal Jeans' mass 
  too high to allow  for baryonic collapse and star formation in 
%formation of 
molecular-cooled haloes in these regions. 
This means that as the reionization proceeds,
 in an increasingly larger fraction of the universe,  baryonic collapse and
star formation is suppressed for  haloes with masses $\la 10^8 \, \rm M_\odot$.
%doesn't allow the 
%formation of these haloes or 
Thus there is a quenching in the formation of new  star forming 
haloes as the reionization proceeds. We take into account
this process in computing the reionization history. (b) It has been argued that the UV light
from the first objects can destroy the molecular hydrogen in the neighbouring haloes, thereby
shutting off the process of further collapse and star formation in these haloes \cite{hrl}. In Sethi, Nath, and Subramanian (2008) \cite{sns}, we showed that in the presence of primordial
magnetic fields, the changed ionized history results in  a sharp increase in the formation
of molecular hydrogen both in the IGM and in the collapsing halo. For magnetic field strength
$B_0 \ge 5 \times 10^{-10} \, \rm G$, the molecular hydrogen in the IGM might be sufficient to block
the UV light from penetrating the IGM (\cite{har00,sns}). (c) Owing to the dissipation of the magnetic field energy in the post recombination era,
the change in the thermal history of the universe can increase the thermal Jeans' 
mass. We take into account this change in our analysis. 

$B_0 \ge  10^{-9} \, \rm G$: In these cases,  the collapsing haloes are in the mass range
which allow both molecular and atomic cooling. This means that the 
impact of quenching, destruction of molecular hydrogen, and the increase
in thermal Jeans' length  discussed above is less pronounced \cite{sns}.

\section{The global HI signal}
The global  HI signal  is observable at the redshifted HI hyperfine transition. The signal
can be seen against the CMBR, the only radio source at high redshifts. The observable 
deviation from the CMBR temperature at any frequency $\nu_0 = 1420/(1+z) \, \rm MHz$ is (e.g. \cite{sr90,mmr,sethi05}):
\begin{equation}
\Delta T_{\rm CMBR} = -{\tau_{\rm \scriptscriptstyle HI} \over (1+z)}(T_{\rm CMBR} -T_s)
\label{eqh1}
\end{equation}
The spin temperature $T_s$ is given by \cite{field58}:
\begin{equation}
T_s = {T_{\rm CMBR} + y_c T_K + y_\alpha T_\alpha \over 1 +  y_c + y_\alpha}
\label{eqts}
\end{equation}
Here $T_K$ is the matter temperature and $T_\alpha$ is the colour temperature of the 
radiation close to the Lyman-$\alpha$ line. In the context of the problem
at hand,  $T_\alpha = T_K$ \cite{field59,rd94}. $y_\alpha$ determines the contribution of Lyman-$\alpha$ radiation in the determination of the spin temperature; $y_\alpha \simeq 0$ before the onset of the epoch of reionization. $\tau_{\rm \scriptscriptstyle HI}$ is given by (e.g. \cite{mmr,sethi05} and
reference therein):
\begin{equation}
\tau_{\rm \scriptscriptstyle HI} = 0.02 \left ({\Omega_{\rm HI} h^2 \over 0.024} \right ) \left ({0.15 \over \Omega_m h^2} \right )^{1/2} \left ({ T_{\rm CMBR} \over T_s} \right ) \left ( {1+z \over 20} \right )^{1/2} 
\label{eqh1p}
\end{equation}
Here $\Omega_{\rm HI}= \rho_{\rm HI}/\rho_c$  is the density 
parameter corresponding  to   the neutral fraction of hydrogen.

The main difference between the standard  zero field  case and the magnetic field model is
 in the thermal
history of the universe. The dissipation of tangled magnetic fields can 
significantly alter
the thermal history of the universe (Paper I). The other 
difference is that the build-up of the Lyman-$\alpha$ radiation  during the 
reionization era  depends on the mass function of 
collapsed haloes, which in our case is computed using 
 the magnetic field-induced matter power spectrum. 
'Lyman-$\alpha$' here
refers to the continuum radiation between Lyman-$\alpha$ and Lyman-$\beta$ frequencies.
The photons in this frequency range escape into the medium without getting 
absorbed by either the emitting source or the HI in the IGM until these photons
redshift to a frequency very close to the Lyman-$\alpha$ resonant line frequency (for 
details e.g. \cite{sethi05} and references therein).
 We model this effect, as 
in Sethi (2005)  \cite{sethi05}, by following the evolution  mean  specific intensity 
of the Lyman-$\alpha$ radiation:
\begin{equation}
I_{\nu_\alpha}(z) \simeq  {H_0^{-1} hc(1+z)^3 \over 4\pi \Omega_m^{1/2}}\int dM\int_z^{z_{\rm max}}dz' {d^2n \over dM dz'}\dot N_{\gamma \alpha}(M,z') (1+z')^{-5/2-\beta} 
\end{equation}
Here $\dot N_{\gamma \alpha}$ is the  Lyman-$\alpha$ photon luminosity  of an
 object and $\beta$ is the spectral index of the photons in the 
frequency range between  Lyman-$\alpha$ and the Lyman-$\beta$. Here
 $1+ z_{\rm max} = (1+z)\nu_\beta/\nu_\alpha$ and is determined from the knowledge 
 that photons  with an energy above the 
Lyman-$\beta$ are absorbed locally (see e.g. \cite{pf06,hirata06}). Given the uncertainty 
about the spectrum of the ionizing sources,  we assume
$\beta = 0$ and $\dot N_{\gamma \alpha} = A \dot N_\gamma$ i.e.
 the 
'Lyman-$\alpha$'  photon luminosity is a constant multiple of the luminosity of 
the  Hydrogen-ionizing 
photons (for more details see the discussion  following Eq.~(10) 
in Sethi 2005 \cite{sethi05}). 

We show in Figure~2 the global HI signal in the post-recombination era
for $A=20$, the ratio of Lyman-$\alpha$ to hydrogen-ionizing luminosity.
The HI signal for one model with zero magnetic field is also shown;
the zero magnetic field case shown corresponds to $A=20$ with 
 spectral index  at energies exceeding 
Lyman-limit   $\alpha=2$ (for further 
details see \cite{sethi05}), and is normalized to the  three year WMAP results. 
%A comparison with
 The figure shows that unlike 
%the HI signal in 
the usual case with zero magnetic fields,
%shows that 
the HI signal  in the magnetised universe is only observable in emission
%in  our case 
throughout the post-recombination era. This is mainly because  the matter
temperature doesn't falls  below the CMBR temperature for the magnetic field models 
we consider (see also \cite{sbk}). 

Different features of Figure~2 can be understood in terms of the altered
thermal history in the presence of the magnetic field. The dissipation of the 
magnetic in the post-recombination era (for details see \cite{ss05}) raises
the matter temperature above the CMBR temperature. As the spin temperature  (Eq.~\ref{eqts})
is determined by  matter temperature at high redshifts, the HI signal 
is observable in emission  at high redshifts. This should be contrasted
with  the standard  case in which  the HI is observable only 
in absorption during the pre-reionization era (Figure~2). 

 As the redshifts decreases, the dilution of the gas density  drives the spin temperature towards CMBR temperature. The HI signal 
therefore approaches zero before the onset of the epoch of reionization. 
As the reionization epoch sets in, the spin temperature gets coupled to 
the  Lyman-$\alpha$ radiation, which results in the spin temperature being driven towards matter temperature
again. In the standard case, depending on the 
details of the onset of reionization, the HI could be observable in 
absorption just  as the reionization begins, as
is seen in Figure~2 (see e.g. \cite{sethi05,gs04}).  However, in the magnetic
field case, the signal is only  observable in emission, as the matter temperature remains above the CMBR temperature. 
This part of the signal is potentially more promising, as it occurs at frequencies more accessible
to the on-going and future experiments. 
The lack of detection of the HI in absorption therefore could signal the 
presence of magnetic fields. 

In both the standard as well the magnetic field case, the HI signal is 
observable in emission during most of the phase of reionization, with the 
signal strength almost independent of spin temperature as the limit $T_s \gg T_{\rm CMBR}$ is reached.

\section{Fluctuating component of HI signal}

Here we analyse this signal in terms of two-point correlation function of the HI signal
in real space. The correlation function can be expressed as (see e.g. 
\cite{zfh04}:
\begin{eqnarray}
\nonumber 
C(r_{12}, \theta) &\equiv& \langle \Delta T(\hat n_1,r_1)\Delta T(\hat
n_2,r_2) \rangle = \\
 & = & T_0(r_1)T_0(r_2) [\langle \psi(\hat n_1,r_1) \psi(\hat n_2,r_2) \rangle - \langle \psi \rangle^2],
\label{corr_fun_f} 
\end{eqnarray}
with
 \begin{equation}
 \psi = \nf({\bf r}) [1+ \delta({\bf r})].
\end{equation}
Here $r_{12} =  |\bf{r_1}-\bf{r_2}|$ and 
$\theta$ is the angle between the line of sight $\hat n$ and the
vector $\bf{r_1}-\bf{r_2}$ separating the two points, and brackets
denote average over  $\bf{r_1}$ (the correlation
function could be written equivalently as a function of the separation angle between
the two points   and the frequency difference $\nu_2-\nu_1$). 
 
The expression in the square brackets in Eq.~(\ref{corr_fun_f}) can be further simplified to (for details see \cite{zfh04,sh08}):
\begin{equation}
\langle \psi \psi \rangle - \langle \psi \rangle^2  =  
 \xi_{xx}
\xi_{\delta\delta}(r_{12}, \theta,z) +
\xi_{xx}  - \bar \nf ^2
\label{corr_fun_psi}
\end{equation}
Here $\xi_{\delta\delta} = \langle \delta({\bf r_1})\delta({\bf r_2})
\rangle$ is the two-point correlation function of the total density
contrast, and $\xi_{xx} = \langle \nf({\bf
r_1}) \nf({\bf r_2}) \rangle$  is the
correlation function owing to the inhomogeneities of the neutral
fraction. For further detail on the assumptions made in deriving Eqs.~(\ref{corr_fun_f})-(\ref{corr_fun_psi}) and the meaning of relevant variables see e.g. \cite{sh08}.
\subsection{Density and ionization correlation functions}
The density correlation function, in the linear regime, can be expressed as (e.g. 
 \cite{ba04,bl05,sh08}) :
\begin{eqnarray}
\nonumber
\xi_{\delta\delta}(r_{12},\theta,z) & \simeq & 
\xi_{\delta\delta}(r_{12},0,z) \left(1 + {2\over 3}\beta + {1\over 5}\beta^2 \right) + \\
&& + \xi_{\delta\delta}(r_{12},2,z) \left({4 \over 3} \beta + {4\over7} \beta^2 \right )P_2(\theta)
\label{corr_fun_rd}
\end{eqnarray}
Here $\beta \simeq \Omega_m^{0.6}/b$, $b$ is the bias of HI distribution with respect
to the matter distribution dominated by the cold dark matter. 
 $P_2(\theta)$ is the Legendre function with $\ell = 2$, and
\begin{eqnarray}
\xi_{\delta\delta}(r,0,z) & = & {D_{+}^2(z)\over 2\pi^2} \int dk k^2 P_{\scriptscriptstyle HI}(k) j_0(kr) \\
\xi_{\delta\delta}(r,2,z) & = & -{D_{+}^2(z)\over 2\pi^2} \int dk k^2 P{\scriptscriptstyle HI}(k) j_2(kr). 
\end{eqnarray}
Here $D_{+}(z)$ is the growing mode of density perturbations and
$P{\scriptscriptstyle HI}(k)$ is the HI  power spectrum; we compute  it as:
$P{\scriptscriptstyle HI}(k) = b^2 P(k)$, where $P(k)$ is the underlying matter power spectrum. Throughout we use $b = 1$. 

As discussed above, we envisage the growth of reionization as expansion and coalescence  of 
spherical ionized bubbles around each source. In this scenario, the two-point correlation function of the 
HI distribution is non-zero even in the absence of density perturbations. The scale of 
ionized inhomogeneities is determined by the scale of the ionized bubbles.

If we assume all these spherical bubbles to be randomly distributed 
and non-overlapping (we will discuss this 
assumption below), then for bubble of  size $R$, one can derive \cite{bl05,sh08}: 
\begin{equation} 
\xi_{xx}   = {\bar \nf^2 \over (1-p_{\rm same}(R))},
\end{equation}
with 
\begin{equation}
p_{\rm same} = (1-\bar \nf)f(r,R).
\label{eq:psame}
  \end{equation}
and 
\begin{equation}
f(r,R) = 1 - {3 r \over 4 R^3}\left(R^2 - {1\over 12}r^2 \right),
\label{eq:frR}
\end{equation}
for $r \le 2R$ and zero otherwise. Here $p_{\rm same}$ is the 
probability that both points in the computation of the two
point correlation function lie in the same ionized bubble. This formalism can be readily extended to bubbles of 
difference sizes if the bubbles are non-overlapping by adding 
contributions from bubbles of different sizes (see e.g. \cite{sh08}):
\begin{equation}
p_{\rm same}(r)   =    \sum_R   g(R) f(r,R)
\end{equation}
Here $g(R)$ is the distribution function of the radii of the HII regions,
normalized such that its integral over all bubble sizes is
the average ionized fraction: $\int g(R) = f_{\rm ion}$.

We show in Figure~3 the distribution function of bubble sizes $g(R)$
at epochs when the average ionized fraction is
$f_{ion}=0.5$ (left panel) and $f_{ion}=0.15$ (right panel) (at redshifts
$z \simeq \{10,12.5\}$, respectively; 
%as in 
 see  Figure~1),
%KS: change OK?
for models with magnetic field values of 
$B_0 = 5 \times 10^{-10} \, \rm G$ (solid curves) and $B_0 = 3 \times 10^{-9} \, \rm G$
(dashed curves).
%, at an epoch when the average ionized fraction is
%$f_{ion}=0.5$ (left panel) and $f_{ion}=0.15$ (right panel).
The dashed-dotted curves correspond to the zero magnetic field case.  

 From Figure~3, we notice the following main differences between the magnetised
and nonmagnetised models:
 (a) the sizes of bubbles
are much smaller in the case of the magnetic field-induced reionization and (b) the
distribution of bubbles radii is far broader  in the standard zero magnetic field case.  
Both these
differences are  due to the difference in the matter power spectrum in the two cases.
As already noted above, most of the haloes are formed close to the magnetic Jeans'
length in the magnetic field case. In the usual case, the power is distributed over
a wider range of scales. Also, as pointed out above,
 the scale of HII regions is smaller in the magnetic field case because
 most haloes are  close to  1-$\sigma$ events and therefore
 the number of collapsed haloes is  larger and the   $\dot N_\gamma$ is smaller, as compared 
to the usual case. 
 An important result of this paper
is that the size distribution of HII regions is almost independent 
of the magnetic field strength as also seen in Figure~3. Even though the 
 mass of the collapsed haloes
 are larger with increasing magnetic field strength, their 
luminosities have to be appropriately lowered to normalize to WMAP results, as already
noted in a previous section. 
 The results shown in Figure~3 for 
the magnetic field models are therefore generic and we note here that we did not
find any exception to these  features  in a wide variety of models we studied. 

\subsection{Effect of Large-scale density field on ionization inhomogeneities}

The results of Figure~3 implicitly assume that the positions of ionizing 
sources are uncorrelated. This however is generally a poor assumption. In the 
usual case, it has been shown that the scale of bubbles is generally much
larger than the assumptions of Figure~3, and the scale of ionization bubble
also evolves very rapidly  (e.g. \cite{bl04,fzh04}). This is also known to be in agreement with 
simulations (e.g. \cite{mellema06,iliev08,mcquinn07}). In the usual $\Lambda CDM$ case,  
the local density  power spectral index in the range of scales of interest to 
the reionization generally lies in the range $-3$ to $-1$, or the density
field is strongly correlated for a large  range of scales. This means that
the impact of the large scale density field on the HII region size is pronounced. Also, the first objects to form are rare objects ($\simeq 2.5 \sigma$), which
means the centers of ionizing sources are more strongly correlated than the 
underlying density field.

In the magnetic field case, the density power spectral index  is $\simeq 1$
above the magnetic Jeans' mass scale, or the density perturbations are strongly suppressed on scales larger than this scale. Also the first objects to collapse are 
close to $1 \sigma$ events which means the centers of the ionizing sources 
are less correlated. This means that the effect of the large scale density
field in this case is likely to be less pronounced. However, we still need
to take it into account to assess its impact. We follow here the prescription
of Furlanetto, Zaldarriaga, and  Hernquist (2004) \cite{fzh04}. 

 Furlanetto et~al. (2004) \cite{fzh04} showed that the comoving number density of the HII
regions with masses in the range $M$ and $M+dM$ can be expressed as:
\begin{equation}
M{dn \over dM} = \sqrt{2 \over \pi} {\rho_b \over M}\left|{dln\sigma \over dlnM}\right|{B_0\over \sigma}\exp\left(-B^2(M,z)/(2\sigma^2)\right).
\end{equation}
Here $B(M,z)=B_0+B_1\sigma^2$ with $B_0=\delta_c(z)-\sqrt{2}K(\zeta)\sigma_{\rm min}$
and $B_1 = K(\zeta)/(\sqrt{2\sigma_{\rm min}})$; $K(\zeta) = {\rm erf}^{-1}(1-\zeta^{-1})$
and $\sigma_{\rm min}$ corresponds to the mass dispersion at the size of the smallest halo
that can cause reionization. We assume this mass to be $M\simeq5\times 10^7 \, \rm M_\odot$ throughout.
The parameter $\zeta$ is the ratio between the ionized fraction and the collapsed fraction at any
given redshift. As $\zeta$ tends to one, the mass function of the HII regions approaches the 
standard Press-Schechter mass function. This is the limit in which 
 the HII regions enclose a single 
ionizing sources, and therefore is expected to approximate the results shown in Figure~3. However, 
$\zeta$ is generally much greater than one, which means the HII regions 
enclose multiple
ionizing sources, and  their sizes evolve far more rapidly than the prescription of Figure~3. 

The size distribution of HII bubbles  is given by the single parameter $\zeta$ in this formulation. 
This parameter would generally depend on various physical parameters e.g. photon luminosity of collapsed haloes, clumping factor, etc, and in general will evolve depending on the history of 
reionization. In other words, the computation of HII regions sizes  would require a physical model of the evolution 
of  the ionized fraction. 
In this paper, we compute this parameter as the ratio of 
ionized fraction to the collapsed fraction at any redshift from the the semi-analytic model described 
in section~2, for the ionization history shown in Figure~1.  

In Figure~4, we show the distribution of HII region sizes for two values of fractional 
ionization, $f_{\rm ion}$ and three values of magnetic field strength. The  values of $\zeta$, in increasing
order of the strength of the magnetic field,  are: $\simeq \{10,14\},\{9,12\},\{4,7\}$, with each
pair of values corresponding to the  ionized fraction $f_{\rm ion} \simeq \{0.15,0.5\}$. These results should be compared to Figure~3. We notice that HII regions sizes are 
larger and evolve more strongly. However, as compared to the usual case (e.g. Figure~4 of \cite{fzh04}), the evolution is not as strong for the 
 reasons discussed earlier
in the section. 

We use the results of Figure~4 to compute the fluctuating  component of the 
HI signal owing to ionization inhomogeneities.

\section{Results: Fluctuating component}

We show the two-point correlation function of the HI fluctuation in Figure~5
 for several
magnetic field models. 
%We also show the expected correlation 
%function, from {\bf purely} density fluctuations (i.e. $\bar \nf ^2 \xi_{\delta\delta}$)
%in the standard case, in the absence of 
%{\bf primordial} magnetic fields. 
All the curves are for $\theta = \pi/2$ (Eq.~(\ref{corr_fun_rd})),
 that is they basically focus on the angular correlation function. 
%The main difference between the two cases is the scales at which the two-point
%correlation function is significant. 
We also break the contribution owing to 
the density and ionization inhomogeneities (the former corresponds to the first 
term in Eq.~(\ref{corr_fun_psi}) and the latter to the next two 
terms in the equation). 

 The first three panels of Figure~5 (counterclockwise from bottom)
correspond to signals which arise for magnetic field values: $\{B_0 = 5 \times 10^{-10}, 10^{-9},
3 \times  10^{-9} \} \, \rm G$, respectively. 
The dashed curves give the contribution of the first term in Eq.~(\ref{corr_fun_psi}),
 that is $\xi_{xx}\xi_{\delta\delta}^B$, 
%magnetically induced density
%perturbations ($\bar \nf ^2 \xi_{\delta\delta}^B$, 
where $\xi_{\delta\delta}^B$ is the correlation function of the purely magnetically 
induced density perturbations. 
The dot-dashed curves show the 
contribution of ionization inhomogeneities to the HI correlation function 
(the $\xi_{xx}  - \bar \nf ^2$ term).
The solid curves in these three panels give the sum of both these contributions.

Note that even in the models with primordial fields, there
is another contribution to HI fluctuations (and correlations) due to
the inflationary induced density perturbations, say referred to as
$\xi_{\delta\delta}^{\rm Inf}$. It should further be pointed out
that if the reionization is caused by primordial magnetic field, 
the photon luminosity of haloes is too small (see discussion on 
Figure~1  above) for the haloes formed owing to
inflationary density perturbation to cause significant contribution.
The number of these  haloes is too small to be important in the 
process of  reionization.  
Therefore, the only impact of the inflationary perturbation in the 
presence of primordial magnetic fields is to add a contribution owing 
to density perturbations. In the fourth panel
(top left panel of Figure~5), we show the contribution 
( $\xi_{xx} \xi_{\delta\delta}^{\rm Inf}$) as a dotted line.
The solid line in the fourth panel
shows the result of adding this contribution as well to
the magnetically induced signals with $B_0 = 3 \times 10^{-9} \, \rm G$.
Thus it shows the predicted total HI correlation function from
all sources of density and ionization fluctuations for the case $B_0 = 3 \times 10^{-9} \, \rm G$.

The results shown in Figure~5 can be summarized as follows:
\begin{itemize}
\item[1.] The typical  scale of ionization inhomogeneities is $\la 1 \, \rm Mpc$ for the 
magnetic field models. This results follows from the nature of distribution function
of HII regions, as shown in Figure~4.
\item[2.] Ionization inhomogeneities  make a comparable contribution to HI fluctuation
signals as that arising from
% to the  contribution from  
density perturbations. For smaller values of magnetic fields,
the ionization inhomogeneities can also provide the dominant contribution of the signal.
\item[3.] The  HI correlation function arising from density fluctuations show
characteristic oscillations, with a scale length which increases  
with increasing value of the magnetic field strength.  This is expected as the dominant 
contribution to the signal comes from scales close to the magnetic Jeans' length.
Detecting such oscillations  could give an indication of 
the influence of primordial magnetic fields
and also help in determining the field strength. Also the change in sign of the HI
correlation at scales above the scale of ionization inhomogeneities ($r \ga 1 \, \rm Mpc$ 
could provide additional information.
%KS: I have changed the text above completely; is it OK?
\item[4.] In Panel 4 of Figure~5 (counter clockwise), we show the 
total observable signal in the presence of a magnetic field of strength 
$3 \times 10^{-9} \, \rm G$.  As noted above  
 if the magnetic field-induced structure formation is
responsible for  reionization,  $\xi_{xx}$ is totally determined by the
primordial field strength, and the only contribution of inflationary
generated density perturbations is to add a term of the form
$\xi_{xx} \xi_{\delta\delta}^{\rm Inf}$ to the HI correlations.

% ionization inhomogeneities in the standard case 
%are  absent.  In other words, the only observable feature of the standard case is the 
%density perturbations modulated by magnetic field induced density and reionization
%inhomogeneities, as shown in Figure~4.  
\end{itemize}

%KS: new text for figure 5 follows:

The main difference between the usual $\Lambda CDM$ case and the 
magnetic field scenario, as also alluded above, is in the scale
of the signal. Both the density coherence scale (as seen in
the fourth panel of Figure~5) and  the scale of ionization
inhomogeneities  are larger in the  $\rm \Lambda CDM$ case.

The process of reionization can be considerably
more complex as compared to the results shown in Figure~1. The ionizing 
sources can have varying spectral indices,  star-formation efficiencies and 
star formation histories, 
which could have a non-trivial impact on the 
photon luminosity  in the process of reionization.
 We have studied a suite of ionization histories in addition to the one shown
in Figure~1 and find that the results in Figure~3 and~4 are fairly
robust.    This behaviour  arises due  to the fact that 
 the magnetic field scenario 
is completely dominated by scales around the magnetic Jeans' length. This 
scale determines both the density inhomogeneities as well as 
the sizes of HII regions that form.

\subsection{Detectability of the signal} 
As noted above,  the main difference between the standard  zero field 
and the magnetic field 
induced reionization is generally the much shorter correlation length scale of the fluctuating HI 
signal in the latter case. 
%On-going and 
Many radio interferometers currently being built to observe the epoch of 
reionization have typical angular resolution of a few arc-minutes ($1' \simeq 2 \, \rm Mpc$ for the cosmological parameters of interest), which corresponds 
to a linear scale $\simeq 5\hbox{--}10 \, \rm  Mpc$. \footnote{These experiments have greater resolution in the 
frequency domain, e.g. MWA can have 4000 channels over 32 MHz band-width,
 which corresponds
to length resolution of $\simeq 15 \, \rm kpc$}. These experiments are
 therefore not 
sensitive to the scales at which magnetic field effects are likely to 
dominate, except for the largest values of $B_0$. MWA is expected to 
have a  primary
beam of $20^0$ and synthesized beam of $4'$ with 500 'tiles'. Assuming full UV 
coverage of the primary beam, the expected error on the estimated two-point
correlation  function
is $\simeq 3 \times 10^{-7}\, \rm K^2$, for an
 integration time of $10^6 \, \rm sec$ (see e.g. \cite{sh08}). This might allow a detection of the 
signal, as shown in Figure~5,  with magnetic field of strength 
 $B_0 \simeq 3 \times 10^{-9}$.
  However, we
note here that an indirect indication of the presence of magnetic field 
might be possible.
%As noted above, in the standard  case, ionization inhomogeneities are likely
 %to play an 
%important role in the detectable signal, as shown in many simulations. 
 %However, 
If the reionization was indeed caused by a 
magnetic field of strength  $B_0 \simeq  10^{-9} \, \rm G$, 
%the only 
%detectable feature would be 
%owing to the density inhomogeneities generated during the inflation era. 
%In other words, 
the reionization process would appear to be  almost 'homogeneous' at 
%these scales. 
the angular resolution probed by these instruments.
%KS: Shiv does the change I have made above look reasonable?

Future radio interferometer SKA has the angular resolution and the frequency
coverage to directly detect the redshifted HI fluctuation owing to magnetic 
fields effects, for the entire range of magnetic field strengths we consider here.
 SKA is projected to have  a point-source 
 sensitivity of $\simeq 400 \, \rm \mu Jy$  in one-minute integration 
(in the continuum mode) in the frequency range of interest ($70 \, \rm MHz$
and $300 \, \rm MHz$).   Assuming a synthesized beam in the range $15''\hbox{--}30''$, relevant for our study, and a simultaneous sky coverage of $25^0$ (the simultaneous sky coverage of SKA is planned to be much larger, which will 
increase the sensitivity further), the sky brightness sensitivity for 
the statistical detection (in the line mode)  is in the range $10^{-6} \, {\rm K^2} \hbox{--} 2 \times 10^{-7} \, \rm K^2$ for one month of integration. This suggests that a 5-$\sigma$ detection of the signal shown in Figure~4 should be possible with less
than one week  of integration with SKA. 

\section{Summary and Conclusions}

Tangled primordial magnetic field can have interesting consequences for 
cosmology. The presence of these fields leave detectable signatures in 
the CMBR temperature and polarization anisotropies (\cite{subramanian3,subramanian4}, \cite{durrer},
\cite{seshadri},\cite{mack}, \cite{gs05},\cite{lewis},\cite{kr05},\cite{gio},\cite{gk08},\cite{yam},\cite{fin08,pao}). The statistics of 
this signature will also have distinctive non-Gaussian features 
\cite{brown,seshadri2,caprini1}. 
%SS: More details about these papers then submitted, astro-ph number?
From these 
considerations, one can obtain an upper limit on the magnetic fields strength 
$B_0 \la 4 \times 10^{-9} \, \rm G$ \cite{gs05}, for models
with  magnetic field power spectrum index $n \simeq -3$, which are
 the only models that are consistent with current state of 
observations \cite{caprini,ss05}. 

Sethi \& Subramanian (Paper I) considered many aspects of these fields in the post-recombination universe. In particular, they showed that magnetic field-induced structure 
formation might lead to early reionization and the dissipation of these fields can 
substantially alter the thermal and ionization history of the universe. 

Magnetic field-induced reionization and its possible impact on the detectable HI 
signal has been considered by several authors \cite{ts06b,sbk}. In this paper,
 we study  these
aspects further within the framework of semi-analytic models which allow us to compute
and add in the impact of reionization inhomogeneities on the HI signal. 

Our main findings can be summarized as:
\begin{itemize}
\item[1.] Owing to magnetic field dissipation in the post-recombination, 
the HI signal is not expected to show any absorption features unlike 
the standard case (Figure~2) (see also \cite{sbk}).
\item[2.] The matter power spectrum owing to magnetic-field induced structure
formation is peaked around the magnetic Jeans' scale.  This scale is imprinted
on the HI signal as typical oscillations in the HI angular correlation function (Figure~5)
\item[3.] Figure~5  also captures the impact of  ionization inhomogeneities.
The scale of the ionization inhomogeneities is 
 $\la 1 \, \rm Mpc$, for a wide range of magnetic field strengths (Figure~4).
 \item[4.] We show that the  
presently-operational and upcoming radio interferometers are sensitive to the 
magnetic field strength $B_0 \simeq 3 \times 10^{-9} \, \rm G$. 
However, the 
future SKA will be able to detect the HI signal for  the entire range of 
magnetic field strength: $5 \times 10^{-10} \la B_0 \la 3 \times 10^{-9} \, \rm G$. 
 A  detection of a characteristic scale and nature of  oscillations
of the HI correlation function  could help measure the magnetic field at 
these scales directly.
\end{itemize}

The upcoming SKA will also have the capability of direct detection of primordial
tangled magnetic field by measuring the Faraday rotation of $\simeq 10^7$
radio sources. 

We note here that the main advantage of a potential detection of the HI signal
is that it is sensitive to far smaller values of magnetic field strength as compared to
the other probes.

\section*{Acknowledgments}
We thank an anonymous referee for many useful suggestions.

\newpage

 \begin{figure}
\epsfig{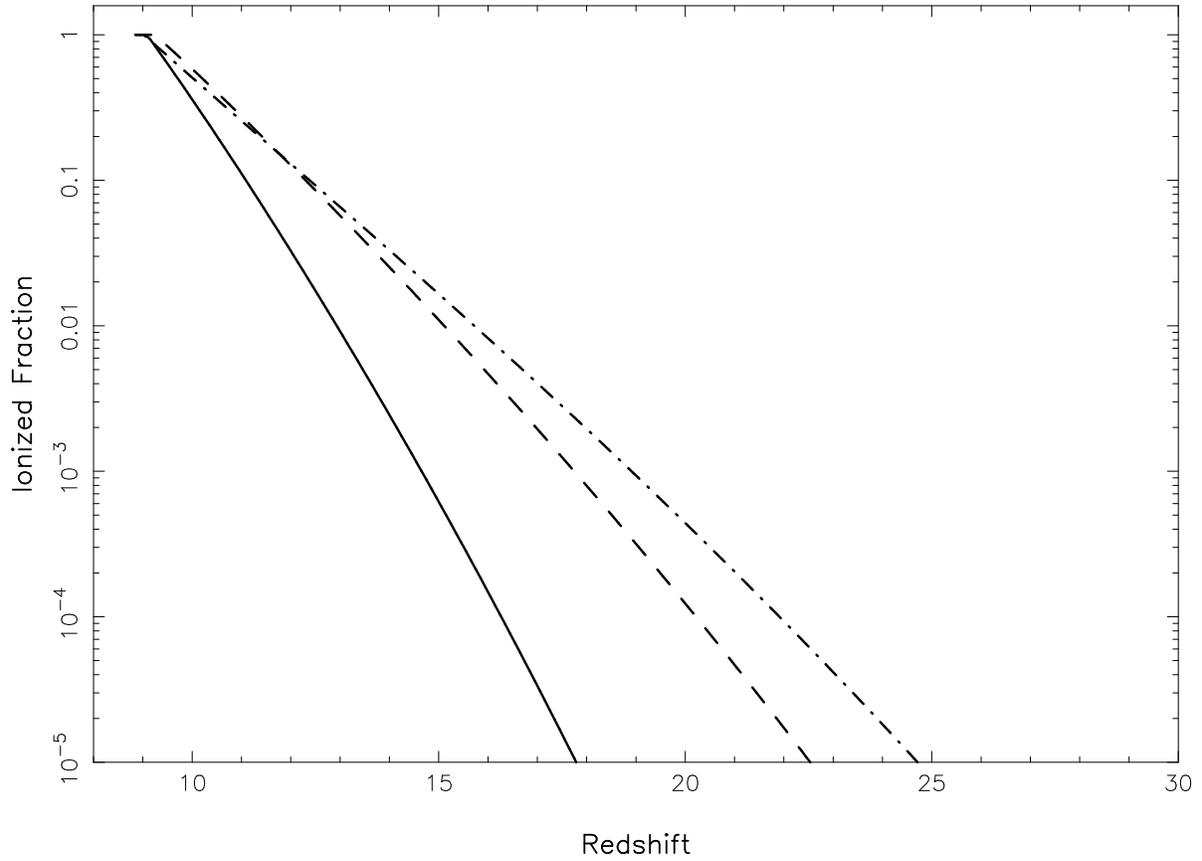}
\caption{The ionization history of the universe is shown for two magnetic field
models along with the usual case without magnetic field. The dashed
and the dot-dashed curve correspond to the magnetic field strength 
$B_0 = \{10^{-9},3\times 10^{-9}\} \, \rm G$, respectively. The solid curve
correspond to the case without magnetic field. For details about other parameters
see the text.}
\label{fig:f1}
\end{figure}

\begin{figure}
\epsfig{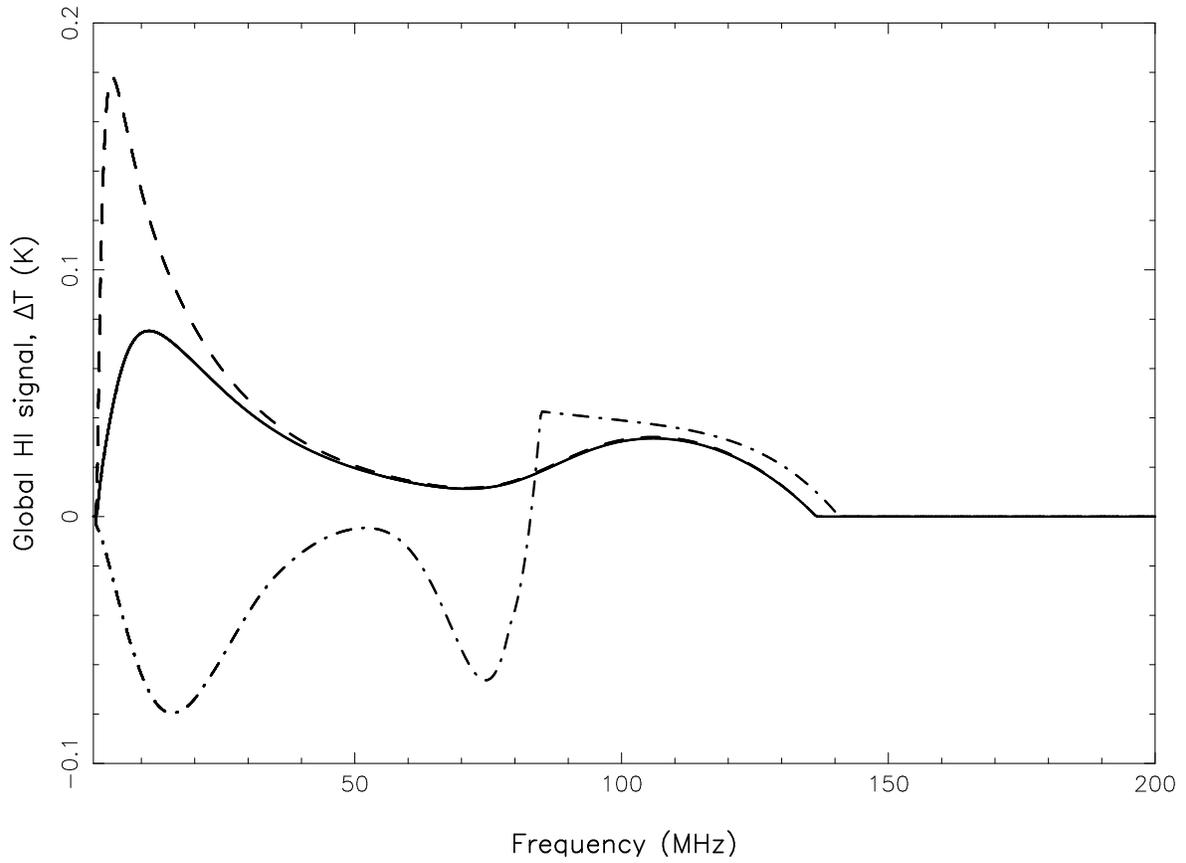}
\caption{The Global HI signal is shown for two values of magnetic field strengths.
The solid and dashed curves correspond to the magnetic field strength 
$B_0 = \{5 \times 10^{-10}, 10^{-9}\} \, \rm G$, respectively. The dot-dashed
curve corresponds to HI signal for one possible  scenario in the 
zero magnetic field case (see text for details). }
%KS: need to put in the 0 magnetic field case also
\label{fig:f2}
\end{figure}

\begin{figure}
\epsfig{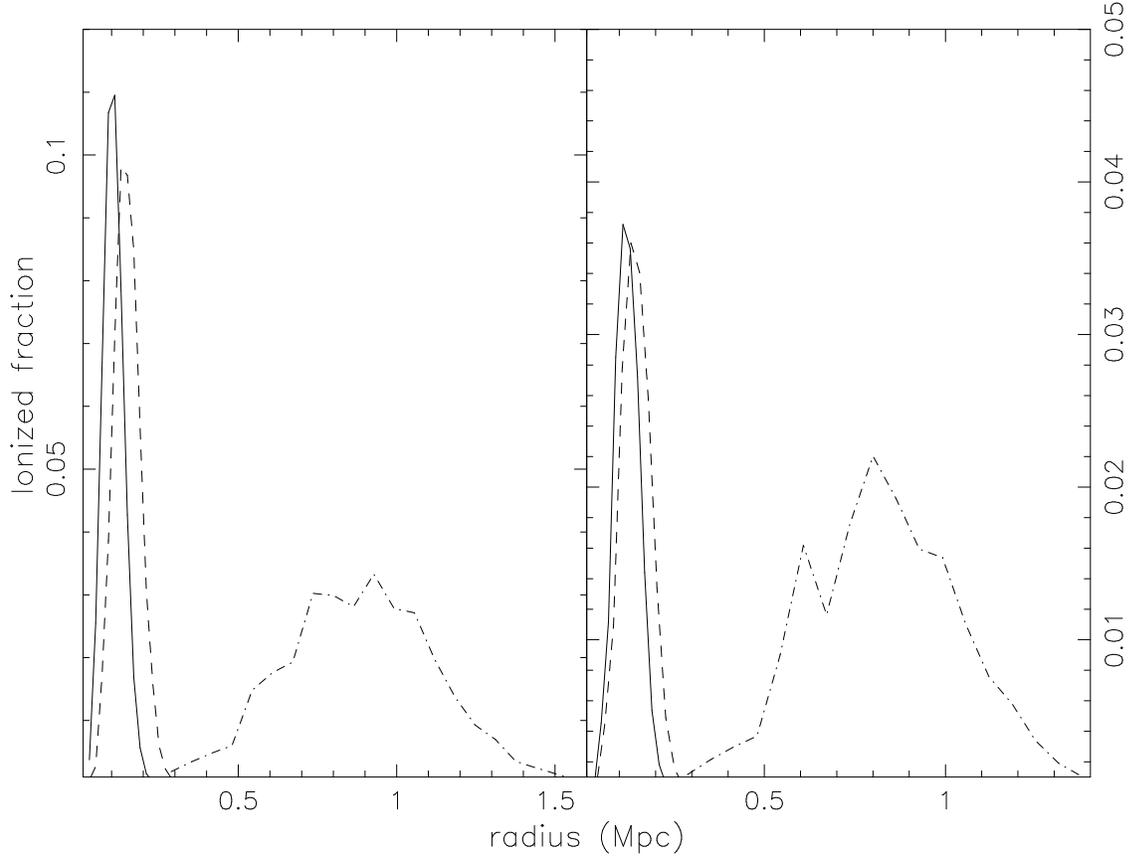}
\caption{The ionized fraction is shown as function of radius of the HII
region. The area under each curves is the ionized fraction. 
The dot-dashed curve correspond to the case without magnetic field.
The solid and dashed  curves correspond to the magnetic field strength 
$B_0 = \{5 \times 10^{-10}, 3 \times 10^{-9}\} \, \rm G$, respectively. 
In the left panel the ionized fraction is $f_{\rm ion} \simeq  0.5$ 
and $f_{\rm ion} \simeq  0.15$ for the right panel, which correspond roughly
to the redshifts $10$ and $12.5$, respectively, as discussed in the text.}
\label{fig:f2}
\end{figure}

\begin{figure}
\epsfig{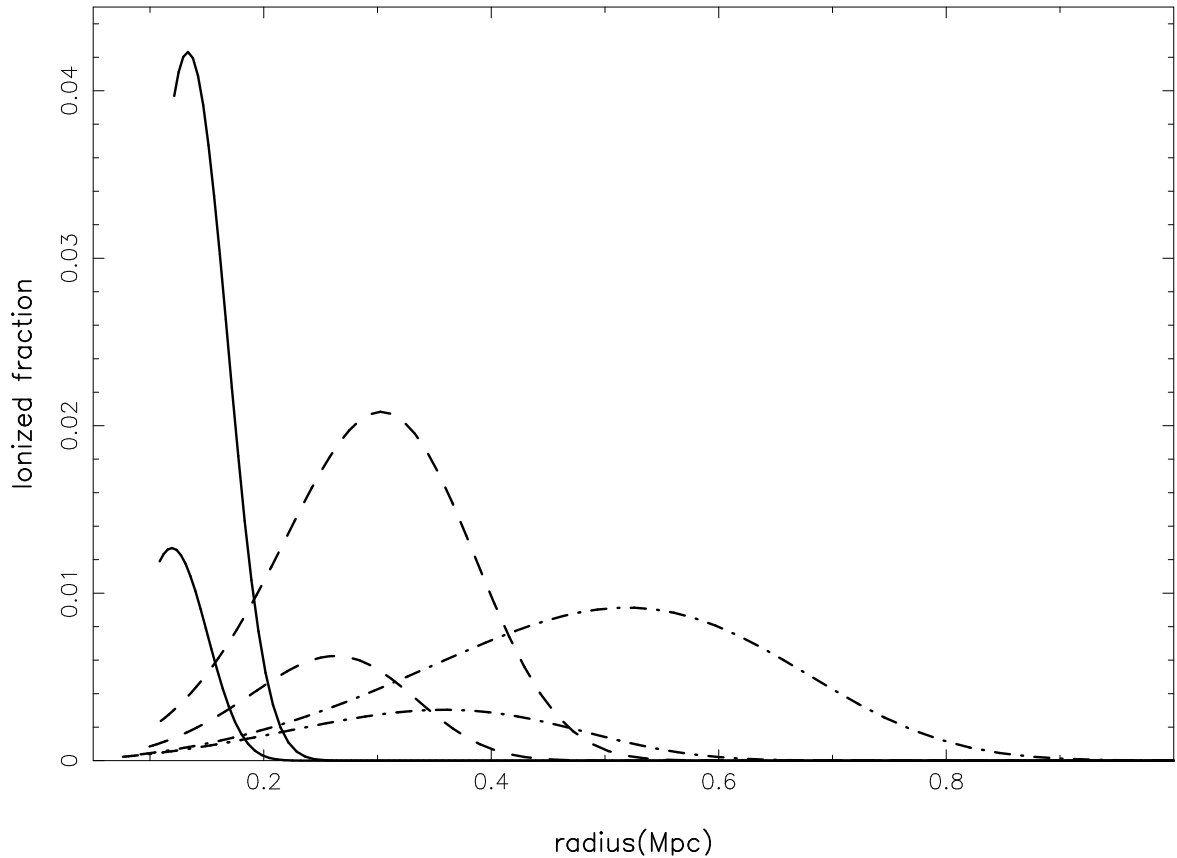}
\caption{The ionized fraction is shown as function of radius of the HII
region.  The solid, dashed, and dot-dashed curves correspond to  $\{B_0 = 5 \times 10^{-10}, 10^{-9}, 
3 \times  10^{-9} \} \, \rm G$, respectively.  The lower and the upper curve for each
line style correspond to $f_{\rm ion} \simeq  0.15$ 
and $f_{\rm ion} \simeq  0.5$, respectively. 
$\zeta$  values for these curves are: $\simeq \{10,14\},\{9,12\},\{4,7\}$, with each
pair of values corresponding to the two cases of ionized fraction. }
\label{fig:f2}
\end{figure}

\begin{figure}
\epsfig{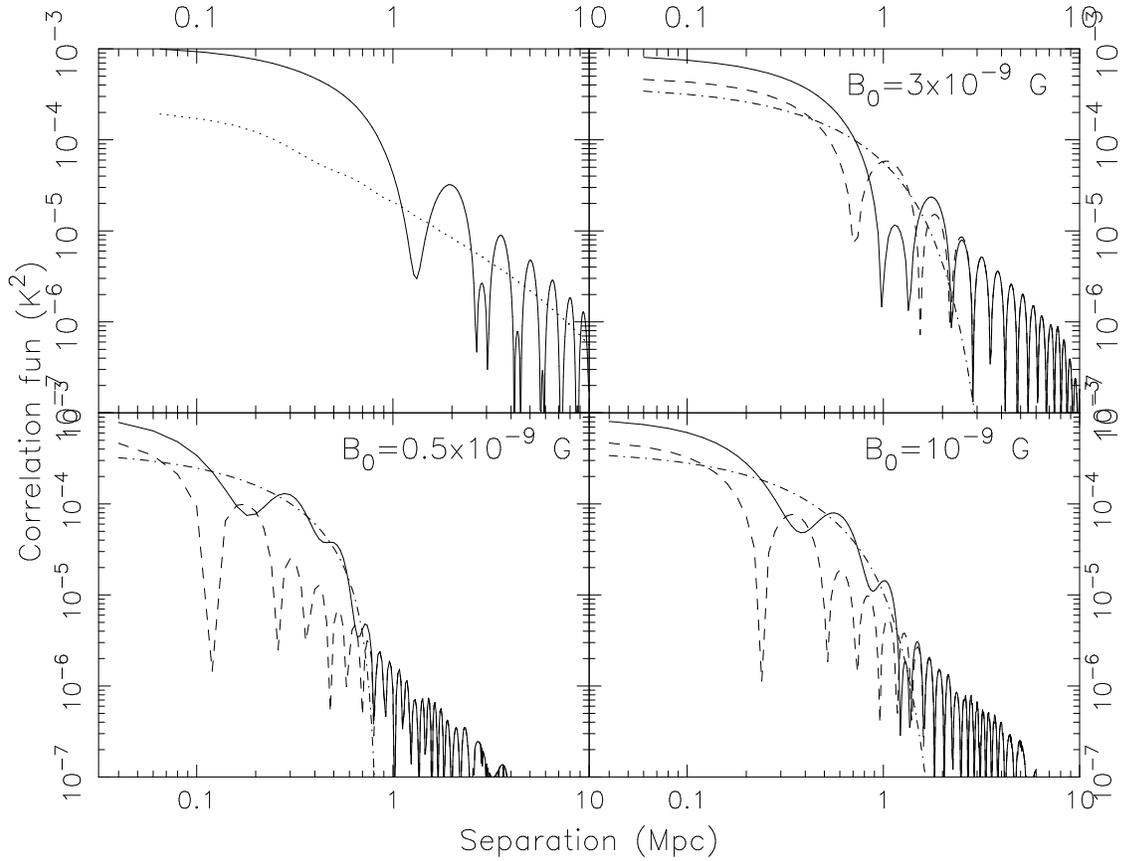}
\caption{The first three panels (counterclockwise from bottom) 
correspond to the signal for magnetic field values: $\{B_0 = 5 \times 10^{-10}, 10^{-9}, 
3 \times  10^{-9} \} \, \rm G$, respectively; The signal   corresponds
to  $f_{\rm ion} \simeq  0.5$ (Left panel of Figure~3) at $z \simeq 10$.  
The solid, dashed, and dot-dashed curves correspond to the absolute
value of the  total signal, the 
signal owing to density fluctuations, and the signal from ionization 
inhomogeneities. In the forth panel, we show the total observable 
signal (solid line) for $B_0 = 3 \times 10^{-9} \, \rm G$, which
 includes the contribution from density fluctuations in the standard 
$\rm \Lambda CDM$ case (dotted line).
%{\bf at what redshifts? Shiv need to rewrite the caption along the
%lines done now in the text, could you have ago? or I will
%do it and send separately later!}
}
%KS: are you now satisfied with the caption???
\label{fig:f4}
\end{figure}

\end{document}